\documentclass{esannV2}

\usepackage[latin1]{inputenc}
\usepackage{hyperref}       % hyperlinks
\usepackage{url}            % simple URL typesetting
\usepackage{booktabs}       % professional-quality tables
\usepackage{amsfonts}       % blackboard math symbols
\usepackage{amssymb}
\usepackage[dvipsnames]{xcolor}
\usepackage{tikz}
\usepackage{bm}
\usepackage{afterpage}
\usepackage{animate}

\usepackage{color, colortbl} %pour mettre de la couleur
\usepackage[super]{nth}
\usepackage{calc}%     needed for the width/height calculations
\usepackage{float}
\usepackage{multirow}

\definecolor{LightGray}{gray}{0.9}

\usepackage{amsmath}
\usepackage{algorithm2e}
\usepackage{algorithmic}
\usepackage{mathtools} % for DeclarePairedDelimiter
\usepackage{bm} %bold math
\usepackage{subfig}
\usepackage{makecell}
\usepackage{amsthm}

\newtheorem*{theorem*}{Theorem}

\newif\ifsubmission
\submissiontrue % you are writing the draft

\newcommand{\eg}{{\em e.g.~}}

\newcommand{\xx}{\bm{x}}
\newcommand{\yy}{\bm{y}}
\newcommand{\ttt}{\bm{\tau}}

\definecolor{colx}{RGB}{255,0,0}
\definecolor{coly}{RGB}{0,120,255}
\definecolor{coltau}{RGB}{1,100,32}

\makeatletter
\newcommand{\gettikzxy}[3]{%
  \tikz@scan@one@point\pgfutil@firstofone#1\relax
  \edef#2{\the\pgf@x}%
  \edef#3{\the\pgf@y}%
}
\makeatother

\DeclarePairedDelimiter\abs{\lvert}{\rvert}%
\DeclarePairedDelimiter\norm{\lVert}{\rVert}%
\makeatletter
\let\oldabs\abs
\def\abs{\@ifstar{\oldabs}{\oldabs*}}
\let\oldnorm\norm
\def\norm{\@ifstar{\oldnorm}{\oldnorm*}}
\makeatother

\usepackage{todonotes}
\newcommand{\marc}[1]{\todo[inline,color=green!40]{#1 -- Marc}}    % marc's notes, inline
\newcommand{\antoine}[1]{\todo[inline,color=blue!40]{#1 -- Antoine}}  
\newcommand{\isabelle}[1]{\todo[inline,color=orange!40]{#1 -- Isabelle}} 
\newcommand{\benjamin}[1]{\todo[inline,color=red!40]{#1 -- Benjamin}} 
\newcommand{\zhengying}[1]{\todo[inline,color=yellow!40]{#1 -- Zhengying}} 
\newcommand{\balthazar}[1]{\todo[inline,color=purple!40]{#1 -- Balthazar}} 

\ifsubmission
    \renewcommand{\marc}[1]{}    
    \renewcommand{\antoine}[1]{}
    \renewcommand{\isabelle}[1]{} 
    \renewcommand{\benjamin}[1]{}
    \renewcommand{\zhengying}[1]{}
    \renewcommand{\balthazar}[1]{}
\fi

\definecolor{mycolor1}{rgb}{1,0.2,0.3}
\definecolor{mycolor2}{rgb}{0.2,0.3,1}
\definecolor{mycolor3}{rgb}{0.,0.4,0.}
\definecolor{coldim1}{rgb}{0.5,0.,0.5}
\definecolor{coldim2}{rgb}{1.,0.5,0.}

\title{LEAP nets for power grid perturbations }

\author{B. Donnot$^{\dagger \ddagger}$\thanks{Benjamin Donnot corresponding authors: benjamin.donnot@inria.com}, B. Donon$^{\dagger \ddagger}$, I. Guyon$^{\bullet \ddagger}$, Z. Liu$^\ddagger$,\\ A. Marot$^\dagger$, P. Panciatici$^\dagger$, M. Schoenauer$^\ddagger$ \\
  $\bullet$ ChaLearn, USA. $\ddagger$ UPSud/Inria, U. Paris-Saclay, France. $\dagger$ RTE France}
  
\begin{document}
\maketitle
\begin{abstract}
We propose a novel neural network embedding approach to model power transmission grids, in which high voltage lines are disconnected and re-connected with one-another from time to time, either accidentally or willfully. We call our architecture LEAP net, for Latent Encoding of Atypical Perturbation.
Our method implements a form of transfer learning, permitting to train on a few source domains, then generalize to new target domains, without learning on any example of that domain. We evaluate the viability of this technique to rapidly assess curative actions that human operators take in emergency situations, using real historical data, from the French high voltage power grid.
\end{abstract}

\begin{figure}[ht!]
\centering
\includegraphics[width=\linewidth]{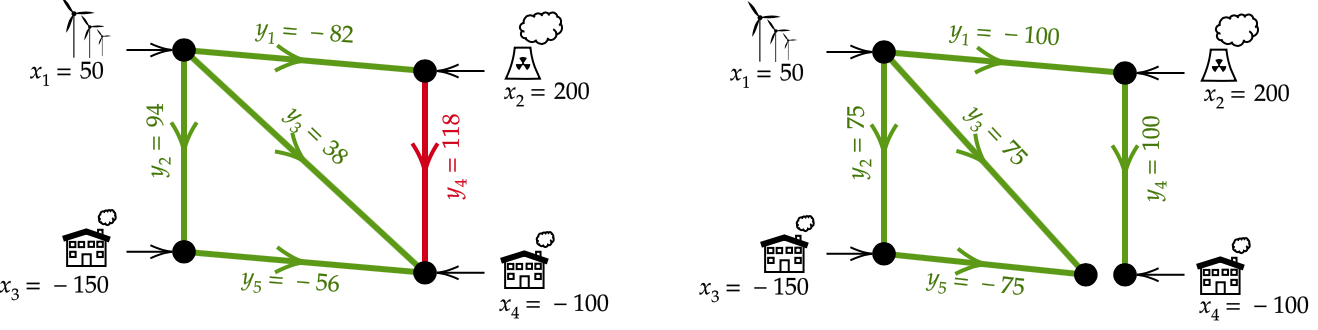} 
\caption{\footnotesize Electricity is transported from production nodes (top) to consumption nodes (bottom), through lines (green and red edges) connected at substations (black circles), forming a transmission {\em grid} of a given topology $\ttt$. Injections $\xx=(x_1, x_2, x_3, x_4)$ (production or consumption) add up to zero. Grid operators (a.k.a. {\em dispatchers}) should maintain current flows $\yy = S(\xx, \ttt)$ below thermal limits. Left: Line $y_4$ goes over its thermal limit 100. Right: A change in topology (splitting of node 6) brings $y_4$ back to its thermal limit.
}

\label{fig:mini_nets}
\vspace{-0.5cm}
\end{figure}

\section{Background and motivations}
We address the problem of accelerating the computation of current flows in power transmission grids, using artificial neural networks, to emulate slower physical simulators, following other pioneering work \cite{Nguyen-95,Hossen-2017,donnot:hal-01695793,donnot:hal-01783669}. Key to our approach is the possibility of simulating the effect of planned coordinated actions on the grid topology (as opposed to accidental suffered changes). Our neural network models may then be used as part of an overall computer-assisted decision process in which human operators (dispatchers) ensure that the power grid is operated in security at all times, namely that the currents flowing in all lines are below certain thresholds (line thermal limits). 
Figure~\ref{fig:mini_nets} illustrates the problem setting on a toy example. If one line goes over its thermal limits, it may be damaged, melt and/or cause fire or break, thus circuit breakers usually put it out of service before this happens. Hence, the grid must be reconfigured quickly to re-balance current flows and avoid that more lines go over their thermal limit, which might result in a cascading effect (black-out).
The space of possible grid topologies grows exponentially with the number of substations. For example, the French high-voltage transmission grid includes $N \approx 6200$ substations, with more than a dozen possible configurations per substation
and thus $\gtrapprox 10^N$ possible grid topologies. Even if only a small number of those are achievable, the search space is still humongous. In practice, Transmission System Operators (TSOs) limit dispatchers to a very limited set of candidate operations. However, operating the grid is becoming increasingly complex because of the advent of less predictable renewable energies, the globalization of energy markets, growth in consumption and concurrent limitations on new line construction. Therefore, it is becoming urgent to optimize more tightly the grid operation, considering a broader range of topological changes operated more frequently, without compromising security.

\section{Proposed methodology}
\label{sec:method}

Our objective is to approximate a function $\yy = S(\xx, \ttt)$ that maps input data $\xx$ (\eg power production and consumption) to output data $\yy$ (\eg power flows), parameterized by a discrete ``grid topology vector'' $\ttt$, taking values in an action space (all possible power-grid topologies \eg  line interconnections). For any fixed topology $\ttt$, training data pairs $\{\xx, \yy\}$ are drawn {\em i.i.d.} according to an unknown probability distribution. In our application setting, $\xx$ is drawn randomly, but $S(\xx, \ttt)$ is a deterministic function implementing Kirchhoff's circuit laws, calculated by a physical simulator that we wish to approximate. 

We call {\bf simple generalization} the capability of a neural net $\hat{\yy} = NN(\xx, \ttt)$ to approximate $\yy = S(\xx, \ttt)$ for test inputs $\xx$ not pertaining to the training set, when  $\ttt$ values are drawn {\em i.i.d.} from a distribution that remains the same in training and test data (this includes the case of a fixed $\ttt$). Conversely, if values of $\ttt$ are drawn according to a {\bf source domain} distribution in training data and from a different {\bf target domain} distribution in test data, then we will talk about {\bf super-generalization}. This setting is a particular case of transfer learning~\cite{pan-yang2010}.

One particularity of our application domain in terms of transfer learning is that we have one {\em primary} ``reference'' source domain (corresponding in the power grid to a reference grid topology $\ttt^{\emptyset} = (0,0,0, \dots)$, around which {\em small} variations are made.
This is a generic scenario in the industry for systems that operate around nominal conditions, thus we anticipate that our method could be extended to other similar situations.  
In our application setting, we can easily get a lot of training data in the reference topology (corresponding to the typical way in which the grid is operated). We have comparably very little data available for training from other {\em secondary} source domains, corresponding to unary changes in grid topology $\ttt^i = (0,0,1, \dots)$ (a single 1 at position $i$). Finally, we have extremely scarce data or no data at all available for training from domains corresponding to double changes $\ttt^{ij}$, or higher order changes (considered target domains). This motivates our architectural design.

    \begin{figure}[ht!]
        \includegraphics[width=1\textwidth]{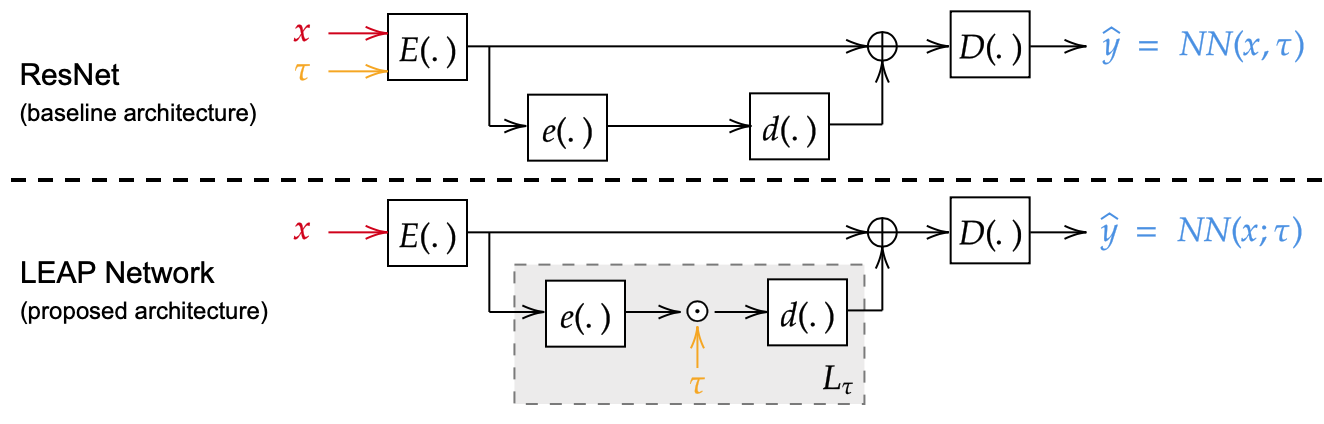}
        \vspace{-0.7cm}
        \caption{ \footnotesize \textbf{Baseline and LEAP architectures:} Top: ResNet~\cite{he2016identity} architecture, with $\ttt$ as input. Bottom: Proposed LEAP net: $\ttt$ intervenes in the latent embedding space. The effect is to make a ``leap'' in latent space.}
        \label{fig:compare_oh_lsi}
        \vspace{-0.5cm}
    \end{figure}
Our proposed Latent Encoding of Atypical Perturbations network, or LEAP net (Figure~\ref{fig:compare_oh_lsi}),  is composed of three parts: An Encoder $\bm{E}$, learning an embedding of the input data $x$; a Decoder $\bm{D}$, learning how to perform the required task within this latent representation; and a Latent module $\bm{L}_{\tau}$, placed between the $\bm{E}$ and $\bm{D}$ where $\ttt$ intervenes. The overall arhitecture is given by:
\vspace{-0.1cm}
    \begin{eqnarray}
        \bm{L}_{\tau} &: & h ~~\rightarrow ~~\bm{d}(\bm{e}(h) \odot \ttt)\\
        \hat{y} &=& \bm{D}\circ (\bm{I} + \bm{L}_{\tau}) \circ \bm{E}(x)
    \end{eqnarray}
\noindent
where $\bm{E}$ and $\bm{e}$ (encoders) and $\bm{D}$ and $\bm{d}$ (decoders) are all differentiable functions (typically implemented as artificial neural networks). The $\odot$ operation denotes the component-wise multiplication and $\circ$ the function composition. If the system is in the reference topology $\ttt^{\emptyset}$, predictions are made according to $\hat{\yy} = \bm{D} \circ \bm{E}(\xx)$. A typical way in which we train LEAP nets is to use a lot of training data in the reference topology $\ttt^{\emptyset}$ (primary source domain), very few examples for each of the unary changes $\ttt^i$ (secondary source domains), and we expect the network to generalize to target domains corresponding to double $\ttt^{ij}$  or higher level changes.

While our architecture draws inspiration from both Dropout~\cite{srivastava2014dropout} and Residual Neural Networks~\cite{he2016identity}, in its mathematical formulation, the underlying concept is quite different. Here we first embed $x$ in a latent space by applying $\bm{E}(x)$. Then, based on $\ttt$ and the location of $\bm{E}(x)$ within the latent space, we compute the corresponding leap $\bm{L}_{\tau} \circ \bm{E}(x)$. Then we decode the signal by applying $\bm{D}$. Those latent leaps contain information about how much the system actually deviates from the reference state, and in which direction. Hence, our architecture only needs to learn to modulate the system response around its nominal value.

\section{Predicting flows in power grids}
\label{sec:powergrid}
We present results for our target application on simulated and real data. Synthetic data allows us to perform controlled systematic experiments and compare neural network approaches with a standard baseline (DC approximation) in power systems. Real data allows us to check whether our method scales computationally while providing prediction accuracies that are acceptable for our application domain.

\subsection{Case 118 synthetic data benchmark}

We conducted controlled experiments on a standard medium-size benchmark from "Matpower" \cite{Zimmerman11matpowersteadystate}, a library commonly used to test power system algorithms \cite{alsac1974optimal}: case$118$, a simplified version of the Californian power grid (dim $\xx$ = $153$ injections and dim $\yy$ = 186 power lines).
Topology changes consist in reconfiguring line connections in one or more substations (see Figure~\ref{fig:mini_nets}). Such changes are more complex than simple line disconnections considered in~\cite{donnot:hal-01695793}.  There are $11~558$ possible unary actions (corresponding to single node splitting or merging, compared to the reference topology). To build the Source domain training and test sets, we sampled randomly $100$ $\ttt^{(i)} \in \mathcal{T}^{Source}$.
In the reference topology ($\ttt^\emptyset$), we sampled $50000$ input vectors $\xx$. But for each $\ttt^{(i)}$, we sampled only $1000$ input vectors $\xx$. 
We used Hades2\footnote{Freeware available at \url{http://www.rte.itesla-pst.org/}.} to compute the flows $\yy$ in all cases.
This resulted in a training set of $150~000$ rows (each row being one triplet $(\xx, \ttt^{(i)}, \yy)$). We created an independent test set of the same size in a similar manner.

We proceeded differently for the Target dataset. We sampled $1500$ (Target domains: $\ttt^{(ij)}\in \mathcal{T}^{Target}$) among the $4950$ possible double actions $\ttt^{(ij)}=\ttt^{(i)} \vee \ttt^{(j)}$, $\ttt^{(i)}$ and $\ttt^{(j)} \in \mathcal{T}^{train}$. Then, for each of these $1500$ $\ttt^{(ij)}$, we sampled $100$ inputs $\xx$ (with the same distribution as the one used for the training and regular test set). We used the same physical simulator to compute the $\yy$ from the $\xx$ and the $\ttt$. The super-generalization set counts then $150~000$ rows, corresponding to $150~000$ different triplets $(\xx, \ttt^{(ij)}, \yy)$.

We compare the proposed LEAP net with two benchmarks: the DC approximation,  a standard baseline in power systems, which is a linearization of the AC (Alternative Current) non-linear powerflow equations, and the baseline neural network architecture (Figure~\ref{fig:compare_oh_lsi}) in which $\ttt$ is simply an input. The mean-square error was optimized using the Tensorflow Adam optimizer. To make the comparison least favorable to LEAP net, all hyper-parameters (learning rates, number of units) were optimized by cross-validation for the baseline network.

\begin{figure}[tb!]
\begin{minipage}[t]{0.48\linewidth}
      \centering
      \subfloat[][Regular gene.]{\includegraphics[width=0.5\linewidth]{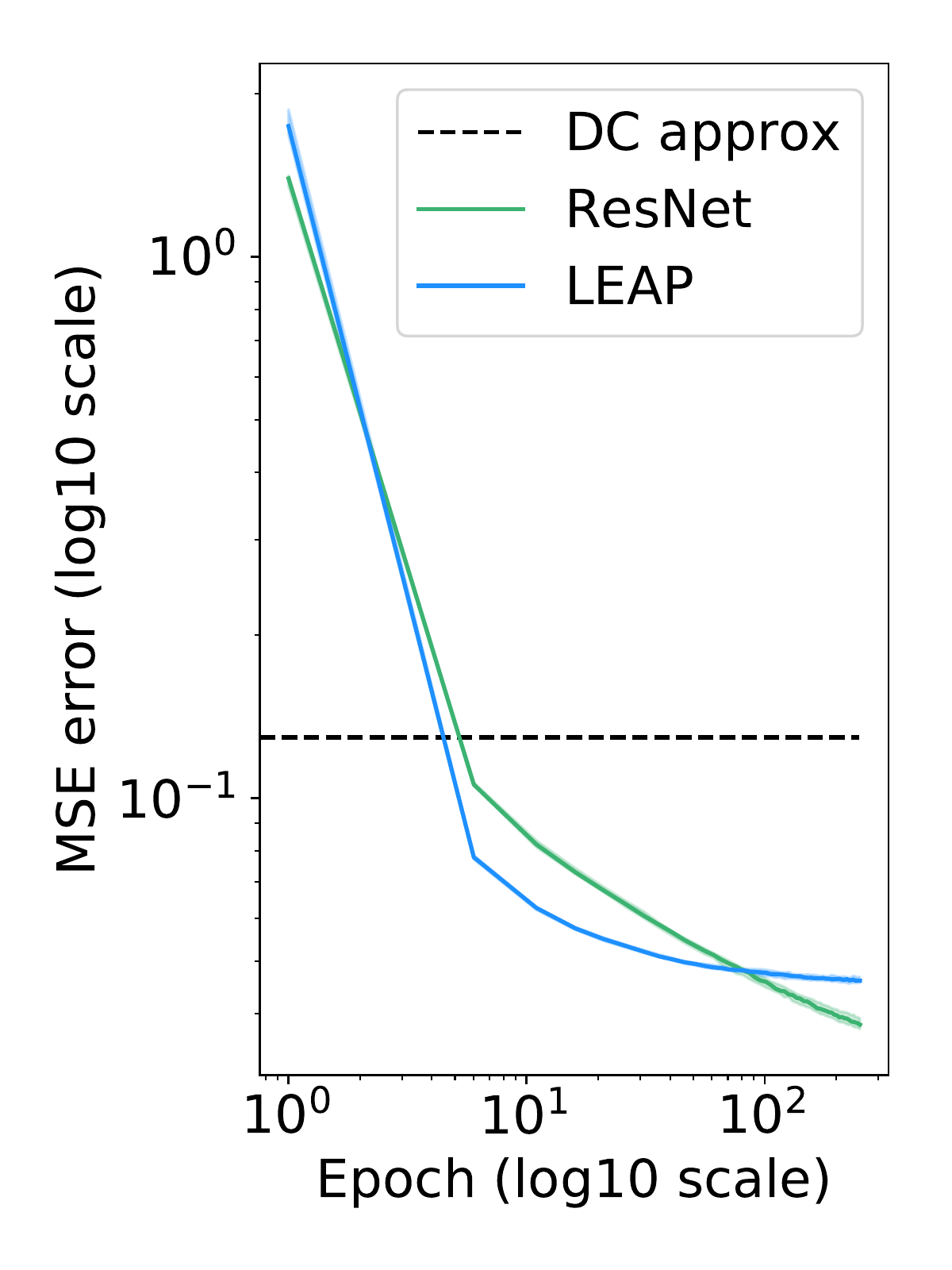} \label{fig:para_test}}
      \subfloat[][Super- gene.]{\includegraphics[width=0.5\linewidth]{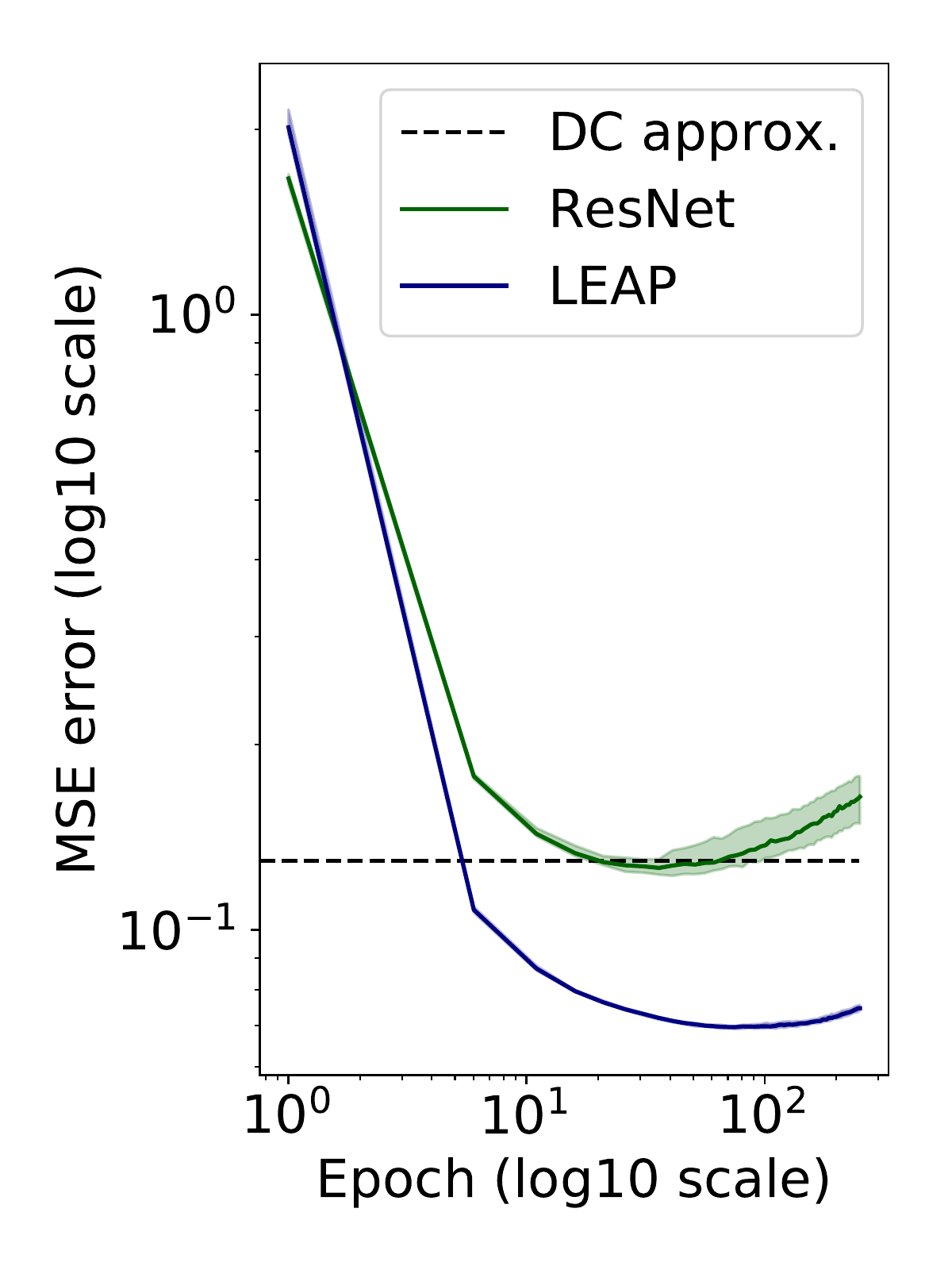} \label{fig:para_gen}}
      \vspace{-0.3cm}
      \caption{ \footnotesize
      {\bf Synthetic data (case 118).} Neural nets trained with 15000 injections, for $\ttt^\emptyset$ and unary changes $\ttt^{(i)}$. (a) {\bf Regular generalization.} Test injections for {\bf unary changes} $\ttt^{(i)}$. (b) {\bf Super-generalization.} Test injections for {\bf double changes} $\ttt^{(ij)}$. Error bars are [20\%, 80\%] intervals, computed over 30 repeat experiments. }
      \label{fig:para}
\end{minipage}
\hfill 
\begin{minipage}[t]{0.48\linewidth}
    \centering
    \subfloat[][Regular gene.]{\includegraphics[width=0.5\linewidth]{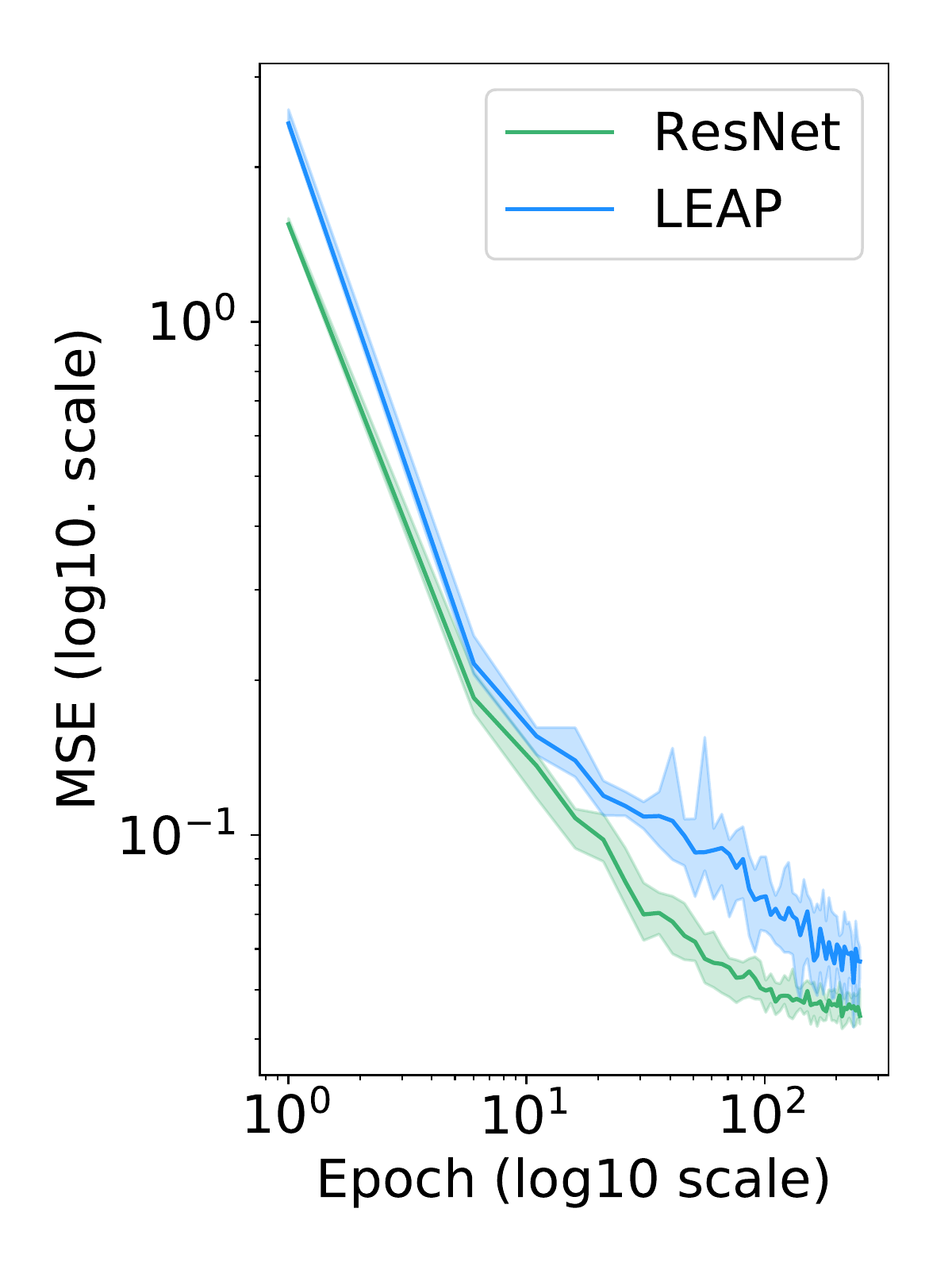} \label{fig:para_toulouse_test} }
    \subfloat[][Super-gene.]{\includegraphics[width=0.5\linewidth]{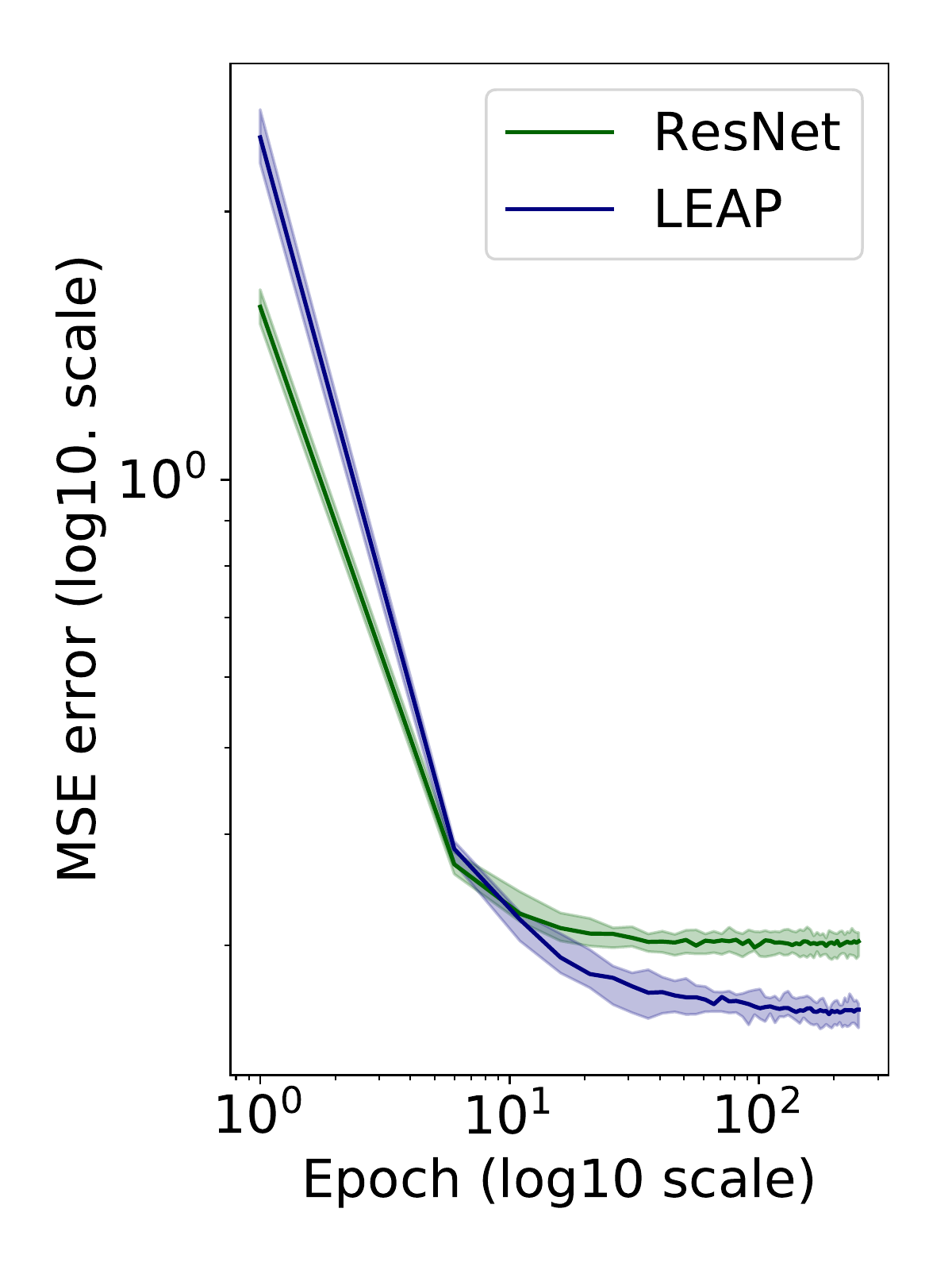} \label{fig:para_toulouse_gen} }
    \vspace{-0.3cm}
    \caption{ \footnotesize
    {\bf Real data from the ultra high voltage power grid.} The neural net in both cases is {\bf trained from data until May 2017}. (a) {\bf Regular generalization.} Test set made of randomly sampled data in same time period as training data. (b) {\bf Super-generalization.} Test set made of the months of June and July 2017. }
    \label{fig:para_toulouse}
\end{minipage}
\vspace{-0.5cm}
\end{figure}

Figure \ref{fig:para} indicates that the LEAP net (blue curves) performs better than the DC approximation (black line) both for regular and super generalization.
Figure~\ref{fig:para_gen} shows that the baseline neural network architecture (green curve) is not viable: not only does it perform worse than the DC approximation, but its variance is quite high. While it is improving in regular generalization with the number of training epochs, its super-generalization performances get worse.
\benjamin{Reviewer 1 (precision):\\

}
\benjamin{Reviewer 3 (hyper parameter):\\
--sizelayerenc=200 --numlayerenc=2 --numlayer=2 \\
1 couche de gd \\
2 unit par topo
}

\subsection{Real French ultra-high voltage power grid data}
We now present results on a part of the French ultra-high voltage power grid: the "Toulouse" area with $246$ consumption nodes, $122$ production nodes, $387$ lines and $192$ substations often split in a variable number of nodes. 
The inputs $\xx$ representing injections (production and consumption) are of dim $\xx = 368$) and the outputs $\yy$ (flows) of dim $\yy = 387$. In this study, $\xx$ and $\yy$  come from real historical data from the company RTE\footnote{Even in real records, flows are estimated, not measured.}. 
One important difference when using played-back data, compared to simulation, is that we cannot intervene (this is strictly observational data). To place ourselves in a realistic transfer learning setting, we used data from 2012 to May 2017 for $\mathcal{T}^{Source}$ and data from June and July 2017 for $\mathcal{T}^{Target}$. This favored changes in $\ttt$ distribution. Another key difference in real data is ``actions space''. In real data {\em actual} grid topologies (specifying line interconnections) are not precisely recorded. Only information on {\em line outages} is available to us as surrogate information on topology. This makes the neural net task harder: it must learn the effects of latent topological changes. This unfortunate loss of information on exact grid topology interventions makes it impossible for us to compare our method to the DC approximation: computing this approximation requires a full description of the topology. The results of Fig. \ref{fig:para_toulouse} yield the same conclusions as in the previous section: the LEAP model generalizes not only to data drawn from a similar distribution it was trained on (Fig. \ref{fig:para_toulouse_test}) but also to unseen grid states (Fig.~\ref{fig:para_toulouse_gen}), better than the reference architecture, which is a critical property for our application.

%%%%%%%%%%%% DISCUSSION %%%%%%%%%%%%

\section{Discussion and conclusion}

The LEAP net architecture has been evaluated on a number of real and artificial test cases.
Training was performed on data triplets $(\xx, \ttt, \yy)$, for which $\ttt \in \mathcal{T}^{Source}$ belong to source domains. The LEAP net generalizes not only by approximating well $\yy$ for new values of $\xx$ when $\ttt \in \mathcal{T}^{Source}$, but also when $\ttt \in \mathcal{T}^{Target}$ (super-generalization). 
In our experiments, we achieved a speed-up of $\approx 300$ times using the LEAP net, compared to running the physical simulator, on the synthetic dataset (power grid of $118$ nodes). With data stored in computer memory, our experiments on the Toulouse area attain a speed of $\simeq 2000$ times compared to running the physical simulator. These computational evaluations were carried out using a single high-end Graphical Processing Unit (GPU) Nvidia Titan X. Further work includes scaling up our method computationally to the entire French extra high voltage power grid. We also need to improve prediction accuracy before our system could be deployed to production. However, the fact that the regular generalization performance is already within an acceptable accuracy range shows great promises. We anticipate several developments. From the theoretical point of view, we could seek mathematical guarantees of super-generalization in the form of performance bounds. It can easily be proved that a LEAP net architecture with linear submodules $\bm{d}$ and $\bm{D}$ exhibits super-generalization with respect to linear superposition of perturbations. However, we have demonstrated experimentally that super-generalization extends to combinations of non-linear perturbations. We are hopeful that more powerful theoretical results could be derived. From the practical point of view, the LEAP net architecture could be used in other application domains, lending themselves to transfer learning.

\bibliographystyle{abbrv}
\bibliography{references.bib} 

\end{document}